\documentclass[aps,preprint,superscriptaddress]{revtex4-1}
\usepackage{color,graphicx}
\usepackage{amsmath}
\usepackage{amsbsy}
\usepackage{amsthm}
\usepackage{bbm}
\usepackage{dcolumn}
\usepackage{bm}
\usepackage{epsfig}
\usepackage{float}
\usepackage[utf8]{inputenc}
\usepackage{color}
\usepackage{natbib}
\usepackage{lipsum}
\usepackage{chemfig}
\usepackage{ulem}
\usepackage{subcaption}
\begin{document}
\title{An investigation into the energy transfer efficiency of a two-pigment photosynthetic system using a macroscopic quantum model}
\author{Fatemeh Ghasemi}
\email{ghasemi\_fatemeh@ch.sharif.edu}
\affiliation{Research Group on Foundations of Quantum Theory and Information, Department of Chemistry,
Sharif University of Technology, P.O.Box 11365-9516, Tehran, Iran}
\author{Afshin Shafiee}
\email{Corresponding Author: shafiee@sharif.edu}
\affiliation{Research Group on Foundations of Quantum Theory and Information, Department of Chemistry,
Sharif University of Technology, P.O.Box 11365-9516, Tehran, Iran}
\begin{abstract}
Despite several different measures of efficiency that are applicable to the photosynthetic systems, a precise degree of efficiency of these systems is not completely determined. Introducing an efficient model for the dynamics of light-harvesting complexes in biological environments is a major purpose in investigating such systems. Here, we investigate the effect of macroscopic quantum behavior of a system of two pigments on the transport phenomena in this system model which interacts with an oscillating environment. We use the second-order perturbation theory to calculate the time-dependent population of excitonic states of a two-dimensional Hamiltonian using a non-master equation approach. Our results demonstrate that the quantum efficiency is robust with respect to the macroscopicity parameter $\tilde{h}$ solely, but the ratio of macroscopicity over the pigment-pigment interaction energy can be considered as a parameter that may control the energy transfer efficiency at a given time. So, the dynamical behavior and the quantum efficiency of the supposed photosynthetic system may be influenced by a change in the macroscopic behavior of the system.
\end{abstract}
\maketitle
\section{Introduction}
Photosynthesis is an interesting kind of energy transformation in nature whereby the solar energy is captured and stored by an organism that converts it into the energy required to proceed life. The absorption of a photon of sunlight by a light-harvesting pigment drives a series of cellular chemical reactions. Pigments, known as light-harvesting complexes (LHCs), are responsible for most of the absorption of sunlight. Light absorption is followed by energy transfer to the reaction center (RC) pigments, where the central electron transfer reactions convert the solar energy into an electrochemical source of energy \cite{1}.
\\In the photosynthesis process, almost all absorbed photons create mobile electronic excited states, called excitons, in the arrangements of antenna chlorophyll (Chl) or bacteriochlorophyll. These excitons are believed to migrate to the photochemically active reaction centers by random walks over the antenna with coherent transport of electron excitation from one Chl to another. Over recent decades, researchers have made significant efforts in monitoring and modeling excitonic energy transfer in molecular systems \cite{2,4,5,6,7,8,9,10,11,12,13,14,15,16,17,18}.
\\Excitation transfer between two pigments can happen in two ways, via Coulomb interaction or quantum tunneling [19]. Although several different measures have been used to quantify the efficiency of natural photosynthesis, the exciton migration and trapping mechanism at the RC is not thoroughly understood. The quantum efficiency is said to be the percentage of absorbed photons leading to stable photosynthetic products and the excitation energy transfer towards the RC occurs with a nearly 100\% quantum efficiency \cite{20,21,22,23}.
\\In recent decades, the nonlinear electronic spectroscopy and theoretical investigations are performed to discover the nature of energy transport across photosynthetic systems. \cite{24,25,26,27,28,29,30}. Engel \textit{et al.} in 2007 using two-dimensional electronic spectroscopy reported that the long-lived quantum coherences between excitonic states play an essential role in the excitation energy transfer in photosynthetic organisms \cite{31}. Their remarkable work proved that the energy transfer in photosynthetic systems is described by coherent wave-like motion, instead of incoherent multi-stage hopping. Observing the long-lasting electronic coherence suggests that quantum coherence might have a significant role in achieving a highly efficient long-range electron flow in photosynthesis \cite{32,33,34}. Long-lasting coherence may overcome the local energetic traps and aid efficient trapping of electronic energy by the pigments covering in the reaction center complex \cite{32}.
\\Consequently, Engel \textit{et al.} in 2011 claimed that the long-lived quantum coherence alone is not sufficient to achieve high quantum efficiency. To this end, the coherences must connect to state populations \cite{35}. In other words, coherence effects must couple to the probability of finding the excitation in a given state. Moreover, the relation between coherence and dynamics of the energy transfer within the photosynthetic systems should be illuminated.
\\To this end, in this study, we consider a two-dimensional Hamiltonian of a two-pigment photosynthetic system, involving trapping terms, to investigate the effect of the quantum macroscopic behavior on exciton transfer between two pigment states in the presence of an oscillating environment. We use the second-order perturbation theory to obtain the time-dependent population of each excitonic state in supposed photosynthetic system. Considering that the pigment-pigment excitation transfer is a macroscopic quantum process, the dimensionless Planck's constant denoted by $\tilde{h}$ determines the extent to which the system has a quantum mechanical behavior \cite{4,8}. In addition, we express the quantum efficiency in terms of the populations of the states. Our results demonstrate that the quantum efficiency is robust concerning the macroscopicity parameter $\tilde{h}$ individually, but the ratio of macroscopicity over the pigment-pigment interaction energy can be considered as a parameter that may govern the energy transfer efficiency at a given time. So, the dynamical behavior and the quantum efficiency for transport phenomena in photosynthetic systems may be influenced by a change in the macroscopic behavior of the system.
\\
This paper is organized as follows. In section II, we first introduce some basic considerations in our formalism then we evaluate the transition rates between two excitonic pigments. Also, we analyze the dynamics of the populations in different situations. In section III, we calculate the quantum efficiency using the probabilities determined in section II. Moreover, we investigate the efficiency of the supposed two-pigment system in different conditions. In section IV, we conclude and briefly discuss the implications of our results from both biophysical and technological perspectives. 
\section{Probability of exciton transfer in a two-pigment photosynthetic system}
\begin{figure}\begin{center}
\includegraphics[trim=150 90 150 70, clip,width=\textwidth]{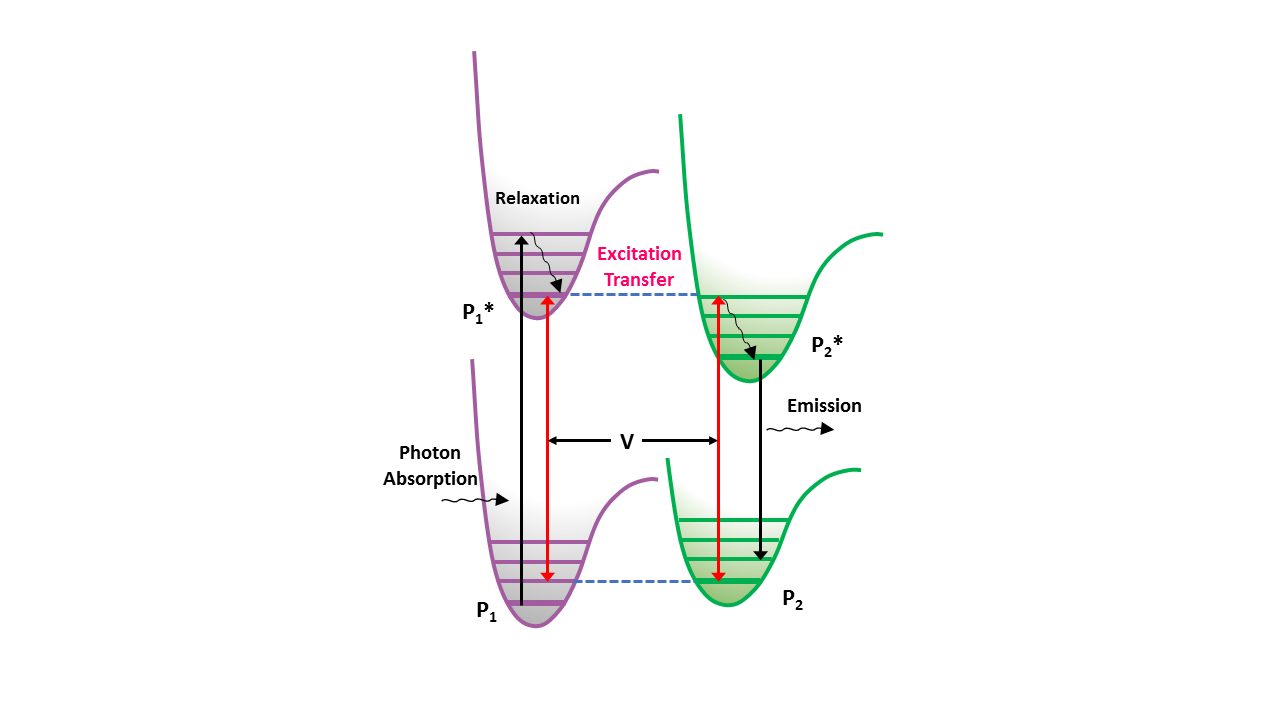}\end{center}

\caption{Energy diagram of photon absorbtion by pigment $P_1$ and the excitation transfer between two pigments $P_1$ and $P_2$.}
\label{Fig1}
\centering
\end{figure}

When the pigment $P_1$ absorbs a photon, one of its electrons is excited to the upper state. Excitation can transfer between two pigments $P_1$ and $P_2$ via Coulomb interaction or quantum tunneling. The exciton transfer between the pigments $P_1$ and $P_2$, for a two-pigment photosynthetic system, can be denoted as ${P_1^*}{P_2}\rightarrow{P_1}{P_2^*}$. Here, we use $^*$ to indicate the excited state. Figure (\ref{Fig1}) shows a schematic diagram of the photon absorbtion and the excitation transfer in a two-pigment photosynthetic system. The dynamics of this process depends on the relative energy of two states and the interaction energies between them. The corresponding Hamiltonian of this process according to quantum mechanics can be given by
\begin{equation}
\label{Eq1}
\hat{H}_s=
\begin{bmatrix}
E_{{P_1^*}{P_2}} & V \\
V & E_{{P_1}{P_2^*}}
\end{bmatrix}
\end{equation}
where $E_{{P_1^*}{P_2}}$ and $E_{{P_1}{P_2^*}}$ are energies of states $\vert{P_1^*}{P_2}\rangle$ and $\vert{P_1}{P_2^*}\rangle$, respectively. This Hamiltonian describes quantum tunneling between the two exciton sites. Moreover, the pigment-pigment interaction is denoted by the Coulomb interaction $V$ which is proportional to $r_{{P_1}{P_2}}^{-3}$, where $r_{{P_1}{P_2}}$ is center-center separation of pigments. This two-pigment system also interacts with the biological environment, including solvent and proteins. We assume the environment as a set of harmonic oscillators with the frequency $\omega_\alpha$, where $\alpha$ runs from $1$ to $N$ in our approach. We modify the Hamiltonian (\ref{Eq1}) to account for the exciton trapping term (we assume that the recombination rates in both sites are equivalent as estimated by Rebentrost \textit{et al.} in \cite{23}) which is defined as
\begin{align}
\label{Eq2}
  H_{\text{trapping}}=\sum_n R_n \vert n\rangle\langle n\vert
\end{align}
\\
where $R_n$ is the trapping rate at the pigment site $n$. The probability that exciton successfully captured at site $n$ during the time interval $dt$ is proportional to the factor $R_n$, given by $2R_n\langle\langle n\vert\rho(t)\vert n\rangle dt$ \cite{33,36}. Thus, the corresponding Hamiltonian matrix for a two-pigment system can be expressed as
\begin{equation}
\label{Eq3}
\hat{H}=
\begin{bmatrix}
E_{{P_1^*}{P_2}}-\tilde{h}R_1 & V \\
V & E_{{P_1}{P_2^*}}-\tilde{h}R_2 \end{bmatrix}
=\begin{bmatrix}e & V \\
V & g
\end{bmatrix}
\end{equation}
where $\tilde{h}$ is the dimensionless Planck's constant characterizing the macroscopicity of the system in the present approach. In the matrix Hamiltonian of the right-hand side we define $e=E_{{P_1^*}{P_2}}-\tilde{h}R_1$ and $g=E_{{P_1}{P_2^*}}-\tilde{h}R_2$. We determine the states $\vert{P_1^*}{P_2}\rangle$ and $\vert{P_1}{P_2^*}\rangle$ in terms of the eigenkets of the Hamiltonian of Eq. (\ref{Eq3}), i.e., $\vert\nu_1\rangle$ and $\vert\nu_2\rangle$ as
\begin{subequations}
\label{Eq4}
\begin{align}
\vert {P_1^*}{P_2} \rangle =\vert e_1 \rangle &=\cos\theta\vert \nu_1\rangle + \sin\theta \vert \nu_2\rangle \\
\vert {P_1}{P_2^*} \rangle =\vert e_2 \rangle &=\sin\theta\vert \nu_1\rangle  - \cos\theta\vert\nu_2 \rangle 
\end{align}
\end{subequations}
where the states $\vert e_1 \rangle$ and $\vert e_2 \rangle$ denote that the excitation is located at site $1$ and $2$, respectively.
We obtain the relation between the angle $\theta$ appearing in Eq. (\ref{Eq4}) and the parameters of the Hamiltonian of Eq. (\ref{Eq3}) as
\begin{equation}
\label{Eq5}
\theta=\dfrac{1}{2}\arcsin(\dfrac{V^2}{V^2+(\dfrac{e-g}{2})^2})^{1/2}
\end{equation}
Also, the eigenvalues of the Hamiltonian (\ref{Eq3}) can be calculated as {\small$E_{\nu_1,\nu_2}=(\dfrac{e-g}{2})\mp[(\dfrac{e-g}{2})^2+V^2]^{1/2}$}. Using the eigenvalues $E_{\nu_1}$ and $E_{\nu_2}$ we difine the tunneling amplitude $\Delta$ as
\begin{equation}\label{Eq6}
\Delta:=\dfrac{\vert E_{\nu_2}-E_{\nu_1}\vert}{\tilde{h}}=\dfrac{2}{\tilde{h}}[(\dfrac{e-g}{2})^2+V^2]^{1/2}
\end{equation}
We investigate the transition between the states $\vert{P_1^*}{P_2}\rangle$ and $\vert{P_1}{P_2^*}\rangle$ by calculating the probability of finding the excitation in each state after time $t$ in the presence of a harmonic environment. The Hamiltonian of the entire system can be introduced as
\begin{equation}
\label{Eq7}
\hat{H}=\hat{H}_s+\hat{H}_\varepsilon+\hat{H}_{s\varepsilon}
\end{equation}
where $\hat{H}_\varepsilon$ is the Hamiltonian of the environment and $H_{s\varepsilon}$ is the system-environment interaction Hamiltonian which in our formalism \cite{38,39,40} has the form
\begin{align}
{\label{Eq8}}
\hat{H}_{s\varepsilon} & = -\sqrt{\dfrac{\tilde{h}}{2}}\sum_\alpha\omega_\alpha^{3/2}f_\alpha(\hat{q})(\hat{b}_\alpha +\hat{b}_\alpha^\dagger)+\dfrac{1}{2}\sum_\alpha\omega_\alpha^2\lbrace f_\alpha(\hat{q})\rbrace^2
\end{align}
where $q$ represents the position variable of the system, $\omega_\alpha$ is the frequency of the environmental oscillators, $\hat{b}_\alpha^\dagger$ and $\hat{b}_\alpha$ are the creation and annihilation operators for the oscillators, respectively, and $f_\alpha(\hat{q})$ describes how the particle $q$ couples to the $\alpha$th environment mode. We use a linearly coupled harmonic environment model, called the separable model, in which $f_\alpha(\hat{q})=\gamma_\alpha f(\hat{q})$, where $ f(\hat{q})$ is an arbitrary function of $q$ and $\gamma_\alpha$ is a positive constant. In Eq. (\ref{Eq8}) all variables are set dimensionless, recall that $\tilde{h}$ is also the dimensionless Planck's constant.
\\We suppose that the initial state of the two-pigment system is $\vert\psi_s(0) \rangle=\vert e_1\rangle=\vert{P_1^*}{P_2}\rangle$ and the environmental initial state is $\vert \text{vac}\rangle$.
Accordingly, the initial state of the entire system is given by
\begin{equation}
\label{Eq9}
\vert\Psi(0) \rangle\rangle=\vert\psi_s(0), \text{vac}\rangle\rangle=\vert\psi_s(0)\rangle\vert\text{vac}\rangle=\vert{P_1^*}{P_2}\rangle\vert \text{vac}\rangle
\end{equation}
In order to investigate the time evolution of the system-environment initial state, we apply the time evolution operator in the interaction picture $\hat{U}_I(t)=e^{-i{\hat{H}_{s\varepsilon}}t/\tilde{h}}$. The time evolution of the initial state $\vert\Psi(0)\rangle\rangle$ can therefore be described as
\begin{equation}
\label{Eq10}
\vert\Psi(t)\rangle\rangle=\hat{U}_I(t)\vert\Psi(0)\rangle\rangle=e^{-i{\hat{H}_{s\varepsilon}}t/\tilde{h}}\vert\Psi(0)\rangle\rangle
\end{equation}
We expand the state $\vert\Psi(t)\rangle\rangle$ in terms of the basis of the direct-product Hilbert space ${{{\mathcal H}_s}} \otimes {{\mathcal H}_\varepsilon}$ as
\begin{equation}
\label{Eq11}
\vert\Psi(t)\rangle\rangle=\sum_{n=0}^{\infty}e^{-iE_nt/\tilde{h}}\vert e_n\rangle\vert \widetilde{\chi _{e_n}(t)}\rangle
\end{equation}
where the states $\vert \widetilde{\chi_{e_n}(t)}\rangle$ are time-dependent coefficients belonging to the Hilbert space of the environment ${\mathcal H}_\varepsilon$, with the following definition
\begin{equation}
\label{Eq12}
\vert \widetilde{\chi_{e_n}(t)}\rangle=\langle e_n\vert\hat{U}_I(t)\vert\Psi(0)\rangle\rangle
\end{equation}
We resort to the perturbation theory, the situation of weak system-environment interaction, to determine the coefficients $\vert \widetilde{\chi_{e_n}(t)}\rangle$. Accordingly, we expand the time-evolution operator $\hat{U}_I(t)$, regarding the interaction Hamiltonian $\hat{H}_{s\varepsilon}$ up to the second order to find
\begin{align}
\label{Eq13}
\hat{U}_I(t) &\simeq  1-\dfrac{i}{\tilde{h}} \int_{0}^{t}\text{d}t_1\hat{H}_{s\varepsilon}(t_1) \nonumber \\
 &-\dfrac{1}{\tilde{h}^2}\int_{0}^{t}\text{d}t_2\int_{0}^{t_2}\text{d}t_1\hat{H}_{s\varepsilon}(t_2)\hat{H}_{s\varepsilon}(t_1)  
\end{align}
where the second and third terms of the right-hand side in Eq. (\ref{Eq13}) are the first and second order perturbative corrections, respectively. Let us now return to the assumption $\vert\Psi(0)\rangle=\vert\psi\rangle\vert \text{vac}\rangle$ and evaluate the expressions $\hat{H}_{s\varepsilon}(t_1) \vert \text{vac}\rangle$ and $\hat{H}_{s\varepsilon}(t_2) \hat{H}_{s\varepsilon}(t_1)\vert \text{vac}\rangle $ to specify the time-dependent coefficients $\vert \widetilde{\chi_{e_n}(t)}\rangle$. When the operator $\hat{U}_I(t)$ applied to the state $\vert\Psi(0)\rangle=\vert\psi\rangle\vert \text{vac}\rangle$, it yeilds following expression
\begin{equation}
\label{Eq14}
\hat{U}_I(t)\vert\text{vac}\rangle \simeq \hat{u}_{\text{vac}}(t)\vert\text{vac}\rangle + \sum_\alpha\hat{u}_{\alpha}(t)\vert\alpha\rangle
\end{equation}
Using Eq. (\ref{Eq14}), one can evaluate the coefficients $\vert \widetilde{\chi_{e_n}(t)}\rangle$ in Eq. (\ref{Eq11}) as
\begin{equation}
\label{Eq15}
\vert \widetilde{\chi_{e_n}(t)}\rangle=\vert\text{vac}\rangle\langle e_n\vert\hat{u}_{\text{vac}}(t)\vert\psi\rangle + \sum_\alpha\vert\alpha\rangle\langle e_n\vert\hat{u}_{\alpha}(t)\vert\psi\rangle
\end{equation}
where the operators $\hat{u}_{\text{vac}}$ and $\hat{u}_{\alpha}$ have the form
\begin{subequations}\label{Eq16}
\begin{align}
\hat{u}_\text{vac}(t) &:= 1-\dfrac{\text{i}}{\tilde{h}}\int_0^t\text{d}t_1\delta\hat{V}(t_1) \nonumber \\ &- \dfrac{1}{2\tilde{h}}\sum_\alpha\int_0^t\text{d}t_2\int_0^{t_2}\text{d}{t_1}\hat{f}_\alpha(t_2)\text{e}^{-\text{i}(t_2-t_1)\omega_\alpha}\hat{f}_\alpha(t_1) \\
\hat{u}_\alpha(t) &:=\dfrac{\text{i}}{\sqrt{2\tilde{h}}}\int_0^t\text{d}t_1\hat{f}_\alpha(t_1)\text{e}^{-\text{i}\omega_\alpha t_1}
\end{align}
\end{subequations}
Using the coefficients $\vert \widetilde{\chi_{e_n}(t)}\rangle$, one can calculate the probability of finding the excitation in the states $\vert{P_1^*}{P_2}\rangle$ ($\vert e_1\rangle$) and $\vert{P_1}{P_2^*}\rangle$ ($\vert e_2\rangle$) at time $t$ as
\begin{subequations}\label{Eq17}
\begin{align}
			& P_{{P_1^*}{P_2}}=\Vert\vert \widetilde{\chi_{{P_1^*}{P_2}}(t)}\rangle\Vert^2 \\
	 		& P_{{P_1}{P_2^*}}=\Vert\vert \widetilde{\chi_{{P_1}{P_2^*}}(t)}\rangle\Vert^2
\end{align}\end{subequations}
In our double-state system the coefficients $\vert \widetilde{\chi_{e_1}(t)}\rangle$ and $\vert \widetilde{\chi_{e_2}(t)}\rangle$, using Eqs. (\ref{Eq4}), (\ref{Eq14}) and (\ref{Eq15}), take the following forms, respectively
\begin{subequations}\label{Eq18}
\begin{align}
\vert \widetilde{\chi_{{P_1^*}{P_2}}(t)}\rangle &=e^{-\text{i}t E_{\nu_1}/h}\cos\theta \langle \nu_1\vert\hat{u}_{\text{vac}}(t)\vert \nu_1\rangle\vert \text{vac}\rangle + e^{-\text{i}t E_{\nu_2}/h}\sin\theta\sum_\alpha\langle \nu_1\vert\hat{u}_\alpha(t)\vert \nu_2\rangle\vert\alpha\rangle \label{18a}
\\ \vert \widetilde{\chi_{{P_1}{P_2^*}}(t)}\rangle  &= e^{-\text{i}t E_{\nu_1}/h}\sin\theta\sum_\alpha\langle \nu_2\vert\hat{u}_\alpha(t)\vert \nu_1\rangle\vert\alpha\rangle - e^{-\text{i}t E_{\nu_2}/h}\cos\theta\langle \nu_2\vert\hat{u}_{\text{vac}}(t)\vert \nu_2\rangle\vert \text{vac}\rangle \label{18b}
\end{align}
\end{subequations}
We see from Eqs. (\ref{Eq17}) and Eq. (\ref{Eq18}) that the problem of calculating $P_{{P_1^*}{P_2}}(t)$ (or $P_{{P_1}{P_2^*}}(t)$) is governed by the matrix elements of the operators $\hat{u}_\text{vac}(t)$ and $\hat{u}_\alpha(t)$.  In this sense, some parity considerations would be useful to realize which matrix elements have non-zero values:
\begin{subequations}\label{Eq19}
\begin{align}
\langle \nu_m\vert\hat{u}_{\text{vac}}\vert \nu_n\rangle= 
\begin{cases} 
      \textit{zero}  &: m-n \text{ is odd} \\
      \textit{non-zero} &: m-n \text{ is even}
   \end{cases}\\
\langle \nu_m\vert\hat{u}_\alpha\vert \nu_n\rangle= 
\begin{cases} 
      \textit{zero}  &: m-n \text{ is even} \\
      \textit{non-zero} &: m-n \text{ is odd}
   \end{cases}
\end{align}
\end{subequations}
Accordingly, we obtain all nonvanishing matrix elements of the operators $\hat{u}_\text{vac}(t)$ and $\hat{u}_\alpha(t)$ as
\begin{subequations} \label{Eq20}
		\begin{align}
			\langle \nu_1 \vert\hat{u}_\text{vac}(t)\vert \nu_1 \rangle &\simeq \text{exp} [ -\dfrac{\text{i}}{\tilde{h}}\lbrace t\delta E_0 - \vert f_{10}\vert^2 F_+(t)\rbrace] \\
			 \langle \nu_2 \vert\hat{u}_\text{vac}(t)\vert \nu_2 \rangle &\simeq \text{exp} [ -\dfrac{\text{i}}{\tilde{h}}\lbrace t\delta E_1 - \vert f_{01}\vert^2 F_-(t)\rbrace] \\                                  
			 \langle \nu_1 \vert\hat{u}_\alpha(t)\vert \nu_2 \rangle &=\langle \nu_2 \vert\hat{u}_\alpha(t)\vert \nu_1 \rangle^*  =\dfrac{2\pi\text{i}}{\sqrt{2\tilde{h}}}\bar{\gamma}_\alpha f_{01}\left( \dfrac{ 1}{\pi}\right) \dfrac{\sin{(\omega +\Delta) t/2}}{\omega +\Delta}e^{\text{i}(\omega + \Delta)t/2}
				\end{align}				
\end{subequations}
where {\small $F_{\pm}(t)=-\pi^{(-1)} \mathcal{P}\int_0^\infty d\omega J(\omega)\dfrac{\sin{(\omega \pm \Delta) t}}{(\omega \pm\Delta)^2}$}. Note that the symbol $\mathcal{P}$ indicates that the integral preceded by it is a principal-value integral and $J(\omega)$, namely the spectral function, has the form $J(\omega):=\dfrac{\pi}{2}\lbrace \bar{\gamma}(\omega)\rbrace^2D(\omega)$.                                                                                                                                                                                                                                                                                                                                                                                                                                                                                                                                                                                                                                                                                                                                                                                                                                                                                  Here, the function $D(\omega)$ represents the frequency distribution of the environmental oscillators and $J(\omega)$ expresses the corresponding distribution weighted by the factor $\lbrace \bar{\gamma}(\omega)\rbrace^2$, describing the strength of the system-environment interaction. In our regime, $D(\omega)$ is defined as $D(\omega):=\dfrac{1}{2\pi t}\lbrace\dfrac{\sin(\omega t/2)}{\omega/2}\rbrace^2$. By evaluating the non-zero matrix elements in Eq. (\ref{Eq19}) and then substituting the results into Eq. (\ref{Eq18}), finally we obtain the probability of finding the excitation in each pigment state at time $t$, as
\begin{subequations} \label{Eq21}
\begin{align}
			& P_{{P_1^*}{P_2}}(t)=\cos^4\theta+\sin^4\theta e^{-\Gamma t}+{2}\cos^2\theta\sin^2\theta\cos t\Delta e^{-\Gamma t/2}\\
	 		 & P_{{P_1}{P_2^*}}(t)=\cos^2\theta\sin^2\theta+\cos^2\theta\sin^2\theta e^{-\Gamma t}-{2}\cos^2\theta\sin^2\theta\cos t\Delta e^{-\Gamma t/2}
\end{align}
\end{subequations}
where the parameter $\Gamma$ is defined as $\Gamma=\dfrac{2}{\tilde{h}}\vert f_{mn} \vert^2 J(\Omega_{nm})$ and represents the dissipation factor of the system-environment interaction. Also, $\Gamma$ represents the strength of the pigment-protein interaction. The parameters $\theta$ and $\Delta$ are defined according to Eqs. (\ref{Eq5}) and (\ref{Eq6}), respectively. Eq. (\ref{Eq21}) represents the time-dependent redistribution of the excitation probabilities from the initial condition. These probabilities are dependent on the pigment-pigment and pigment-protein interaction characteristics such as $\theta$, $\Delta$ and $\Gamma$.
\begin{figure*}[t!]
\begin{subfigure}{0.3\textwidth}
  \centering
 \includegraphics[trim=50 40 10 60, width=0.9\linewidth]{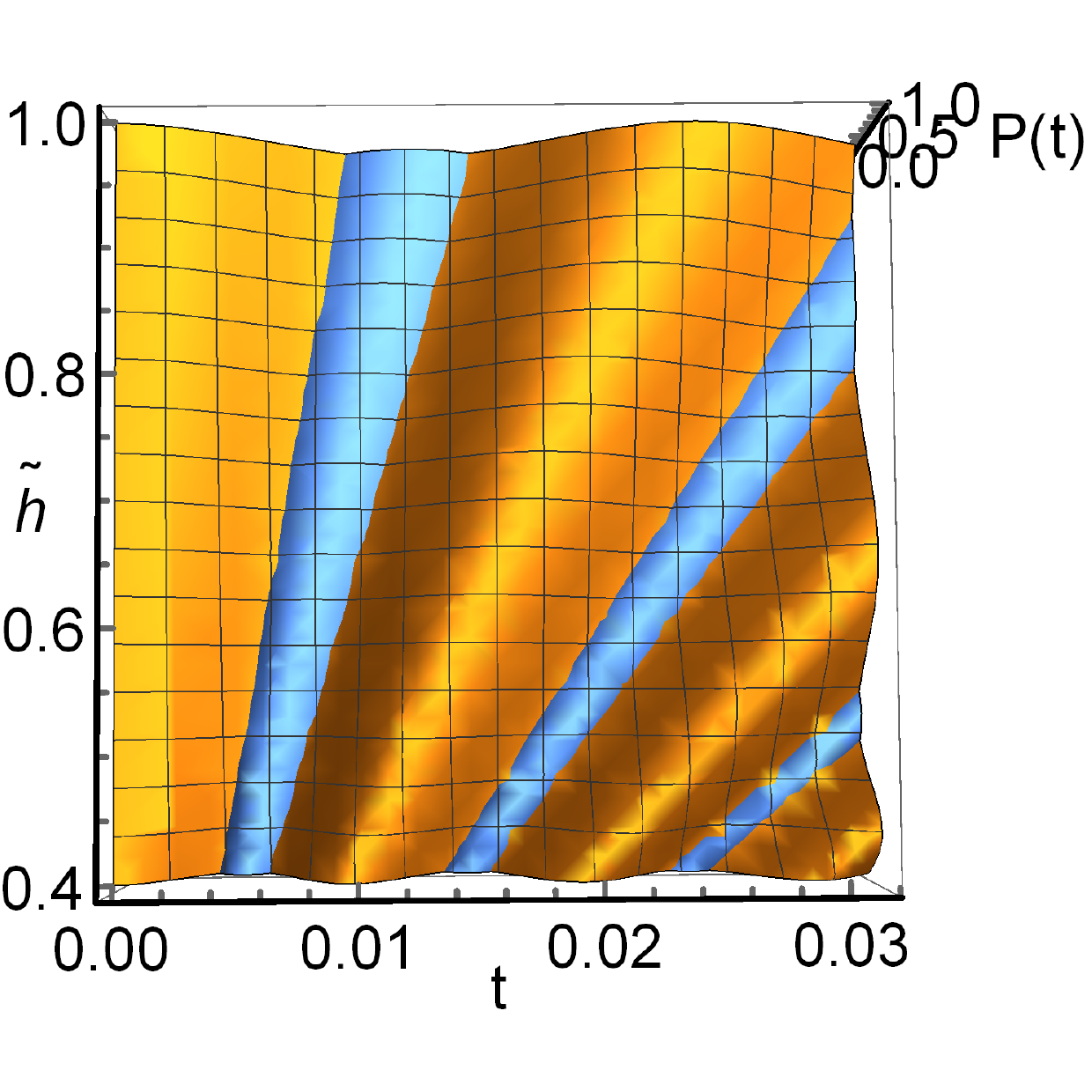}
 \caption{}
 \label{fig2(a)}
\end{subfigure}
~
\begin{subfigure}{0.3\textwidth}
  \centering
 \includegraphics[trim=50 40 10 60, width=0.9\linewidth]{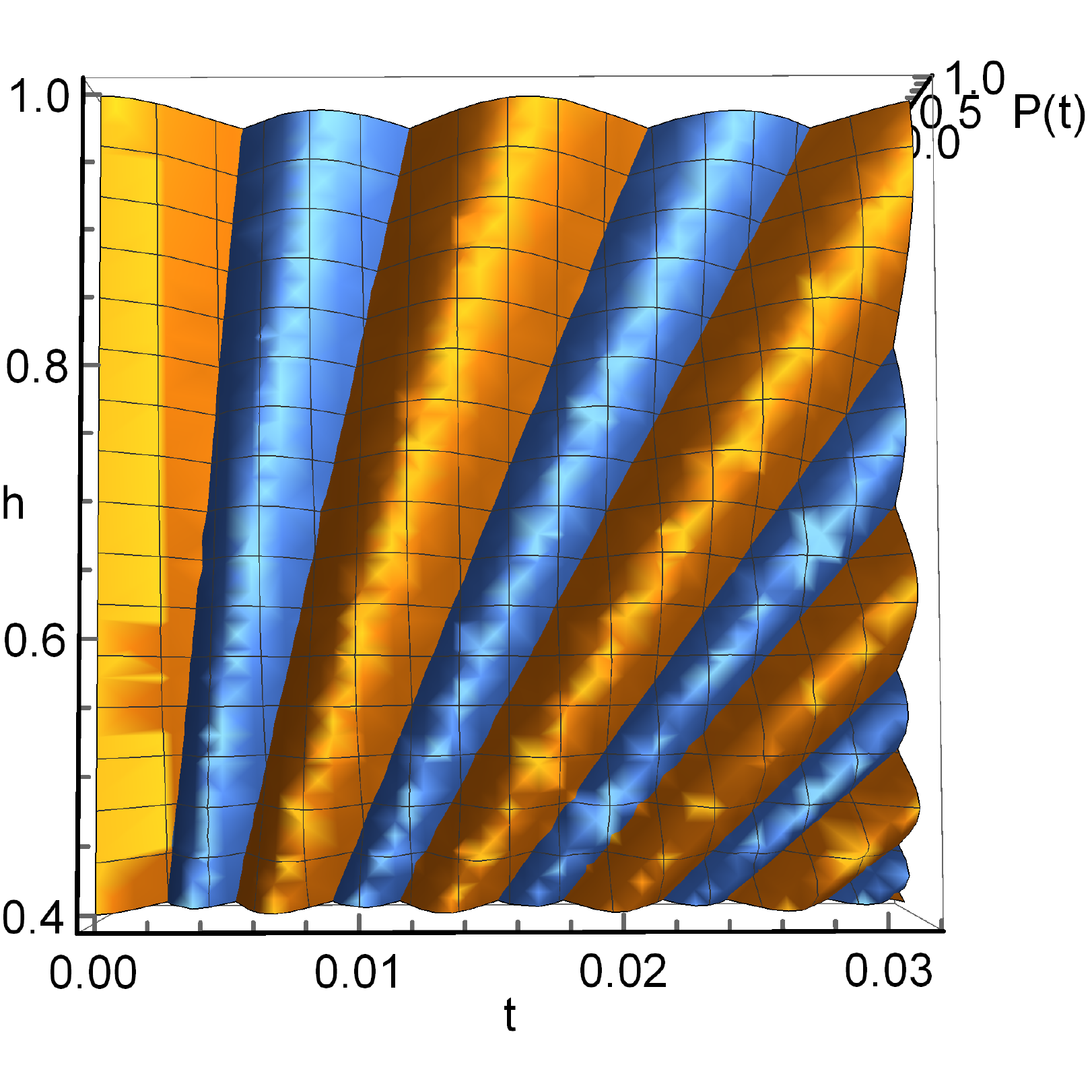}
 \caption{}
 \label{fig2(b)}
\end{subfigure}
~
\begin{subfigure}{0.3\textwidth}
  \centering
 \includegraphics[trim=50 40 10 60, width=0.9\linewidth]{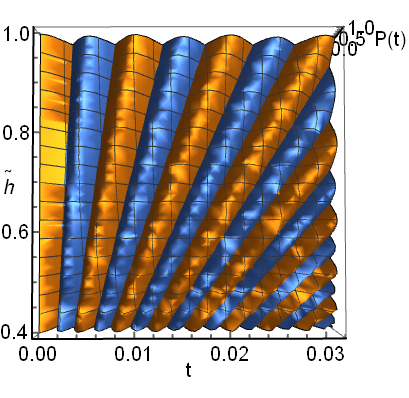}
 \caption{}
 \label{fig2(c)}
\end{subfigure}
\caption{Top views of 3D plots of probabilities. (a) Plot of $P_{{P_1^*}{P_2}}$ (\textit{blue}) and $P_{{P_1}{P_2^*}}$ (\textit{orange}) vs $\tilde{h}$ and $t$ with the values $V=130$ cm$^{-1}$, $e-g=230$ cm$^{-1}$ and $\Gamma=0.01\Delta$, (b) Same as (a) but with $V=230$ cm$^{-1}$, (c) Same as (a) but with $V=398$ cm$^{-1}$.}
\label{Fig2}
\end{figure*}
\\Figure (\ref{Fig2}) shows the variation of the probability of finding excitation in the system of two sites $P_1$ and $P_2$. Fig. (\ref{fig2(a)}) to Fig. (\ref{fig2(c)}) demonstrate that populations how may be affected versus the variation of $\tilde{h}$ and $t$, for different constant value of pigment-pigment interaction. For more detailed information, we have plotted the probabilities as the function of time $t$ in Fig. (\ref{Fig3}). According to Fig. (\ref{Fig3}), the site populations could be strongly affected by pigmen-pigment interactions and therefore by the geometry and arrangement of pigments. As the interplay between two sites is not very strong, the excitation is more localized to the site $P_1$. On the other side, as the pigments lie in the closer distance the exciton will be delocalized in both sites with more similar probabilities. In addition Fig. (\ref{Fig3}) shows that the population fluctuate more fast for large values of $\tilde{h}$ in Figs. (\ref{fig3(d)}) to Fig. (\ref{fig3(f)}), which $\tilde{h}=1$ with respect to Figs. (\ref{fig3(a)}) to Fig. (\ref{fig3(c)}) which $\tilde{h}=0.5$. However, the averege amplitudes are not affected by changing $\tilde{h}$. Another side view of the Fig. (\ref{Fig2}) is demonstrated in Fig. (\ref{Fig4}), which shows the variation of the exciton site probabilities vs the macroscopic trait of the two-pigment system. Figs. (\ref{fig4(a)}) to Fig. (\ref{fig4(c)}) show that the interaction energy $V$ determines the amplitude of the population fluctuations. It is worth to note that we found that the magnitude of $\Gamma$ determines the decay rate of probabilities.
\begin{figure*}[t!]
\begin{subfigure}{0.3\textwidth}
  \centering
 \includegraphics[trim=50 10 10 40, width=0.9\linewidth]{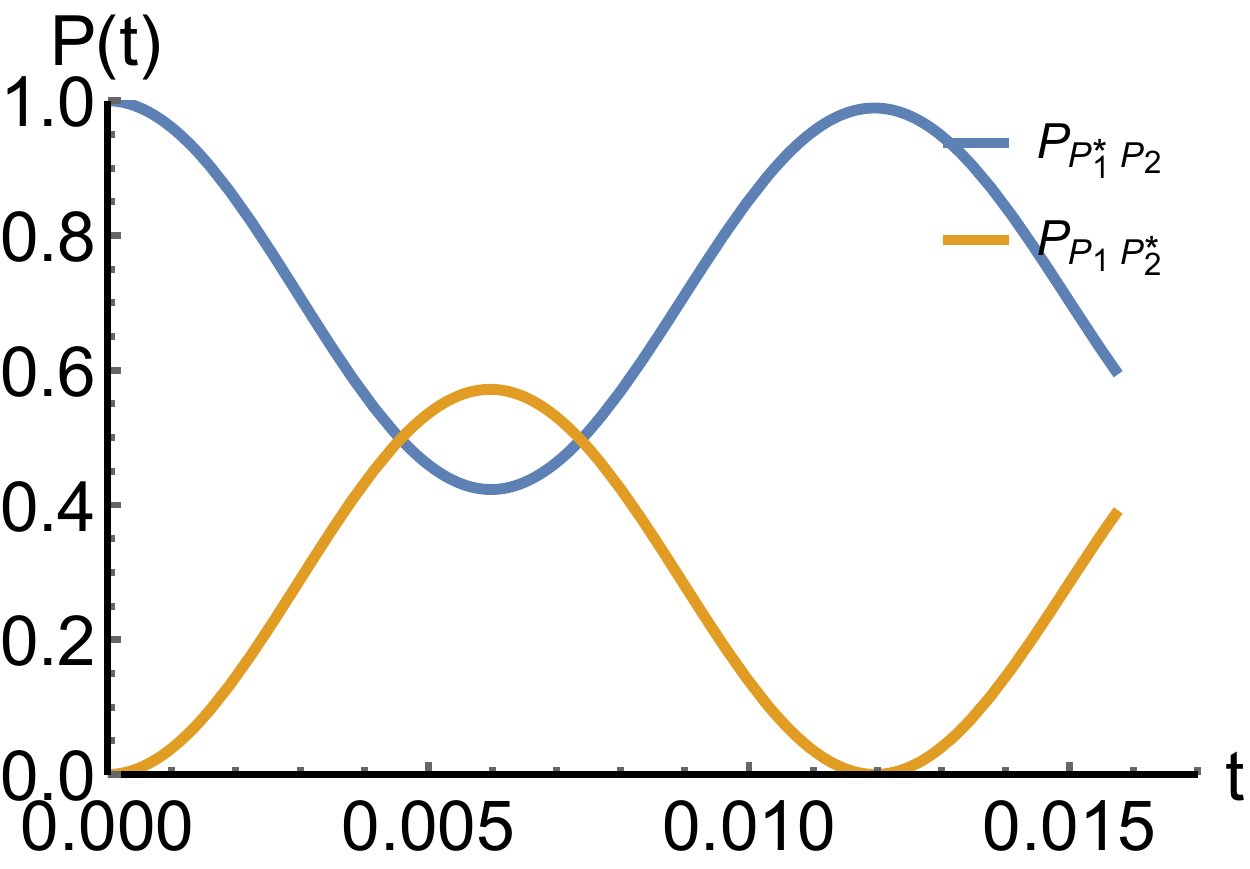}
 \caption{}
 \label{fig3(a)}
\end{subfigure}
~
\begin{subfigure}{0.3\textwidth}
  \centering
 \includegraphics[trim=50 10 10 40, width=0.9\linewidth]{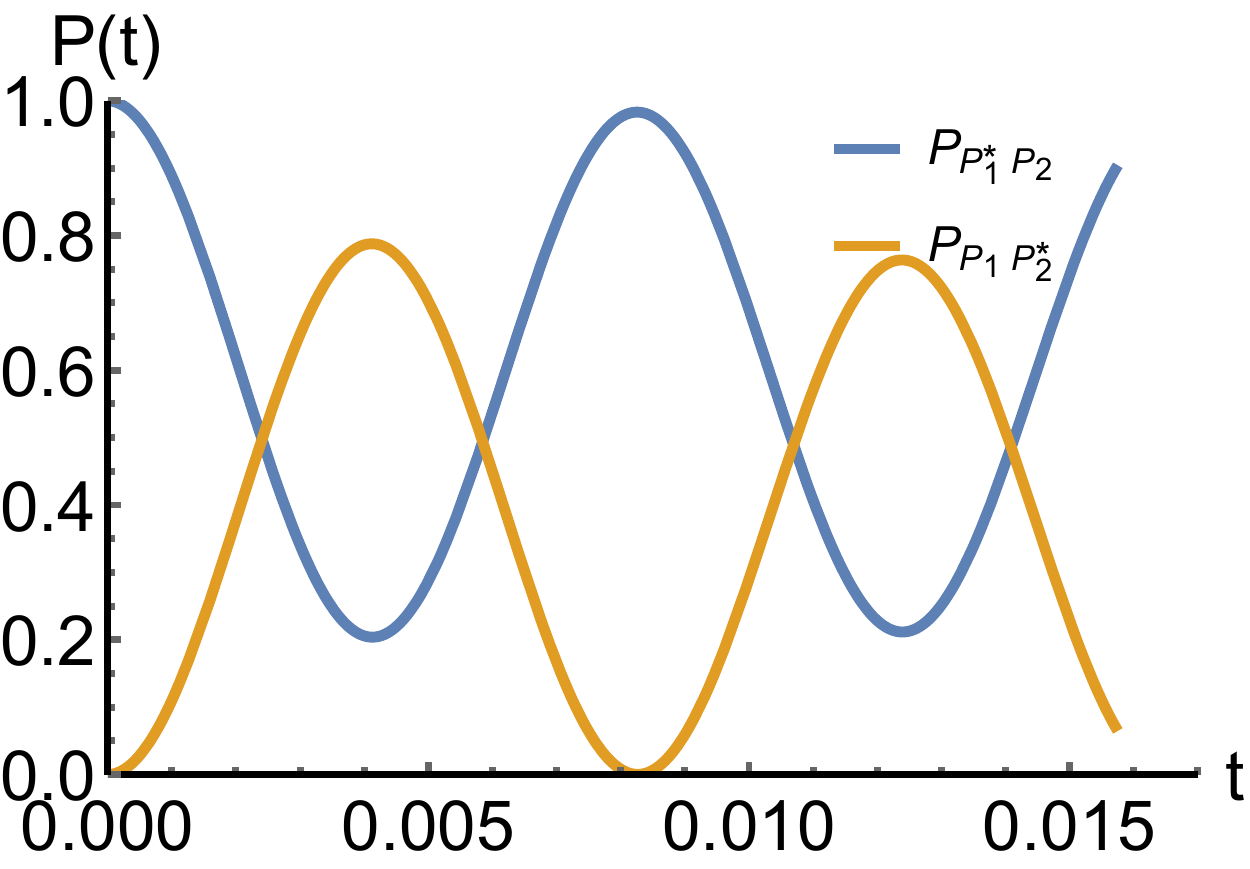}
 \caption{}
 \label{fig3(b)}
\end{subfigure}
~
\begin{subfigure}{0.3\textwidth}
  \centering
 \includegraphics[trim=50 10 10 40, width=0.9\linewidth]{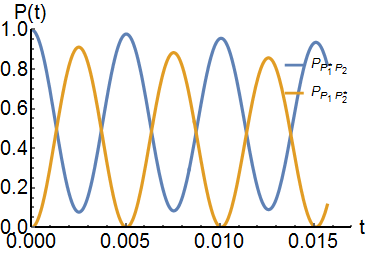}
 \caption{}
 \label{fig3(c)}
\end{subfigure}
~
\begin{subfigure}{0.3\textwidth}
  \centering
 \includegraphics[trim=50 10 10 10, width=0.9\linewidth]{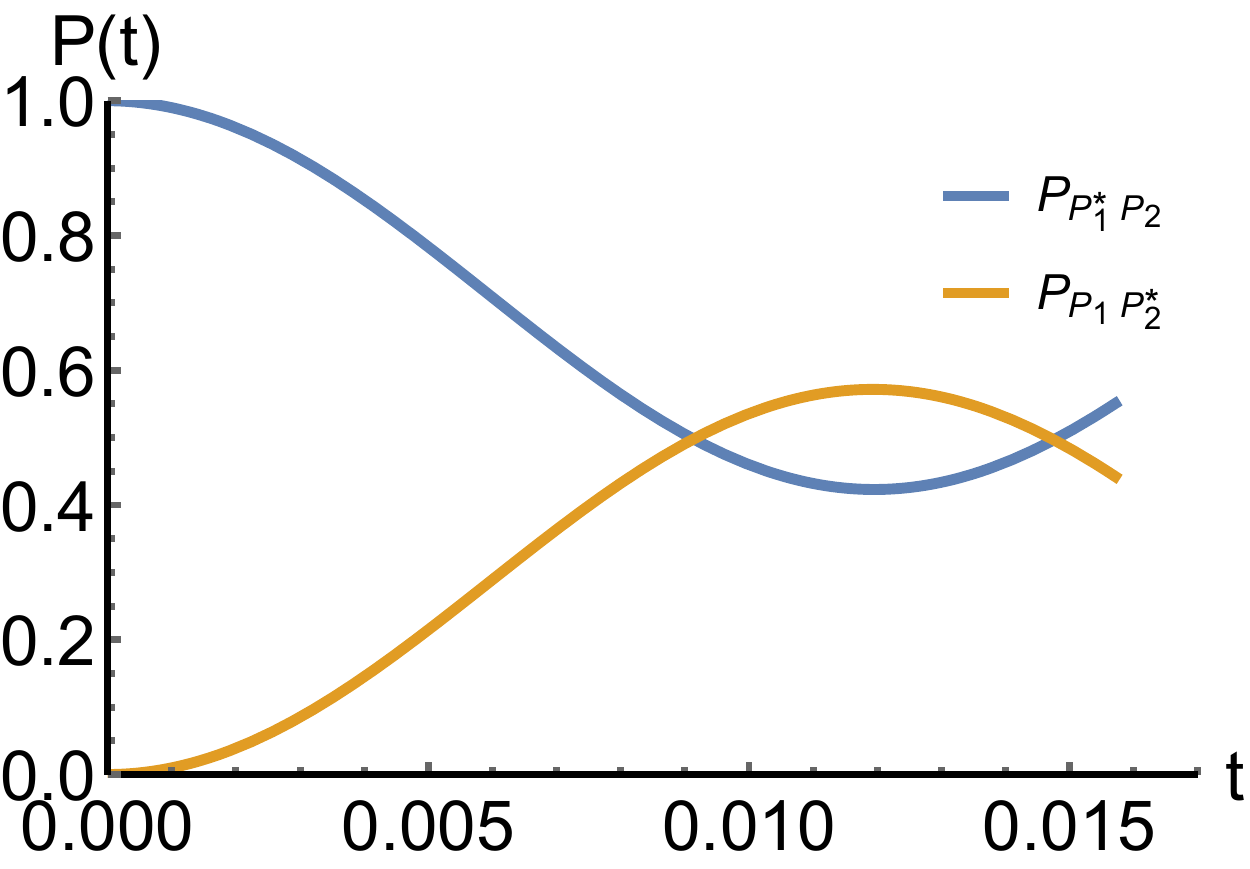}
 \caption{}
 \label{fig3(d)}
\end{subfigure}
~
\begin{subfigure}{0.3\textwidth}
  \centering
 \includegraphics[trim=50 10 10 10, width=0.9\linewidth]{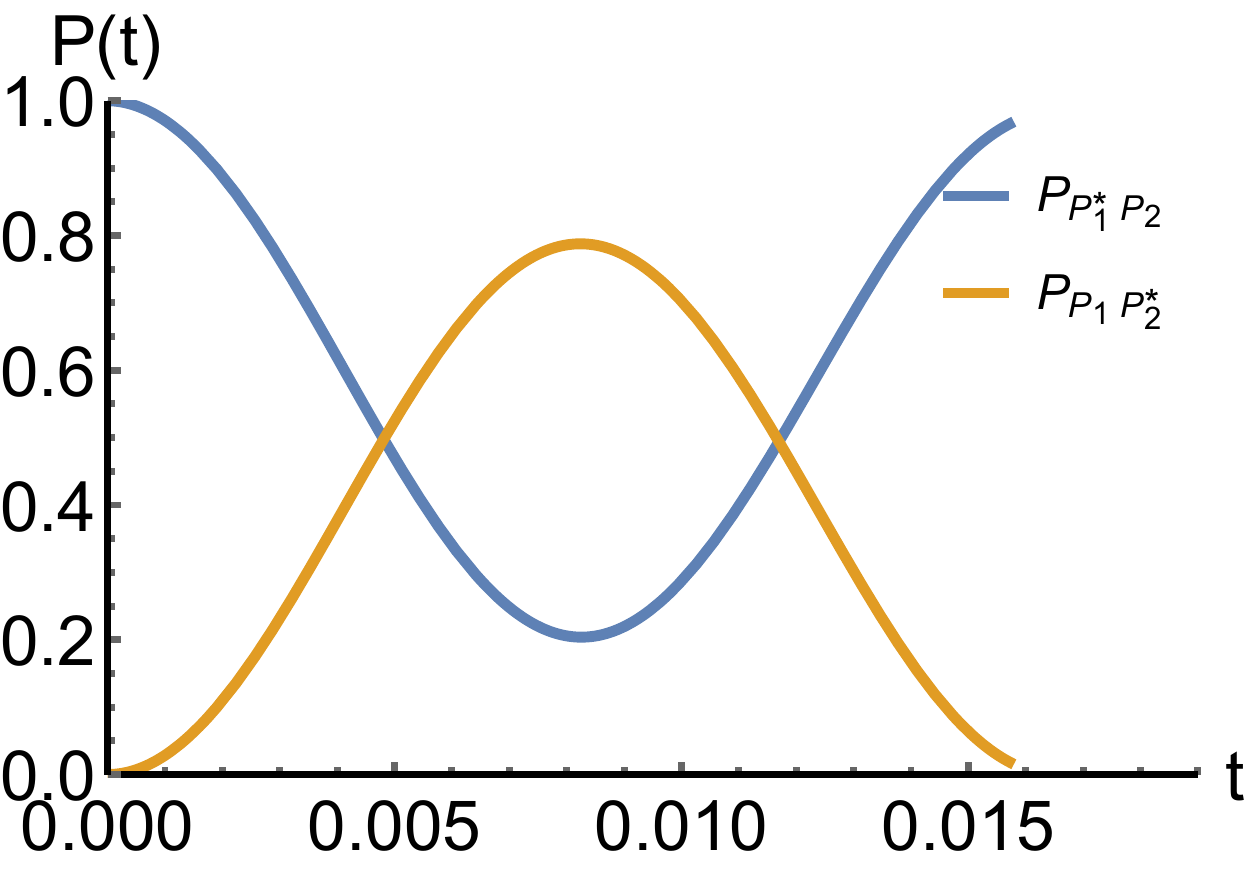}
 \caption{}
 \label{fig3(e)}
\end{subfigure}
~
\begin{subfigure}{0.3\textwidth}
  \centering
 \includegraphics[trim=50 10 10 10, width=0.9\linewidth]{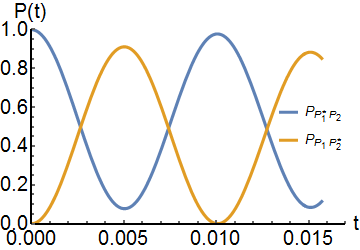}
 \caption{}
 \label{fig3(f)}
\end{subfigure}
\caption{Variation of the probability of finding excitation in the pigments $P_1$ (\textit{blue}) and $P_2$ (\textit{orange}), (a) plot of $P_{{P_1^*}{P_2}}$ and $P_{{P_1}{P_2^*}}$ as a function of $t$ with the values $V=130$ cm$^{-1}$, $e-g=230$ cm$^{-1}$, $\tilde{h}=0.5$ and $\Gamma=0.01\Delta$ within the principal time domain, (b) Same as (a) but with $V=230$ cm$^{-1}$, (c) Same as (a) but with $V=390$ cm$^{-1}$, (d) plot of $P_{{P_1^*}{P_2}}$ and $P_{{P_1}{P_2^*}}$ as a function of $t$ with the values $V=130$ cm$^{-1}$, $e-g=230$ cm$^{-1}$, $\tilde{h}=1$ and $\Gamma=0.01\Delta$, (e) Same as (d) but with $V=230$ cm$^{-1}$, (f) Same as (d) but with $V=390$ cm$^{-1}$.}
\label{Fig3}
\end{figure*}
\\
\begin{figure*}[t!]
\begin{subfigure}{0.3\textwidth}
  \centering
 \includegraphics[trim=50 1 10 1, width=0.9\linewidth]{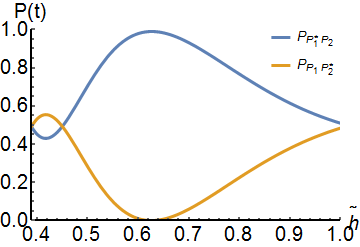}
 \caption{}
 \label{fig4(a)}
\end{subfigure}
~
\begin{subfigure}{0.3\textwidth}
  \centering
 \includegraphics[trim=50 1 10 1, width=0.9\linewidth]{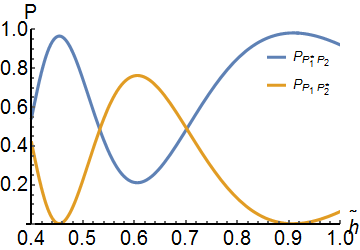}
 \caption{}
 \label{fig4(b)}
\end{subfigure}
~
\begin{subfigure}{0.3\textwidth}
  \centering
 \includegraphics[trim=50 1 10 1, width=0.9\linewidth]{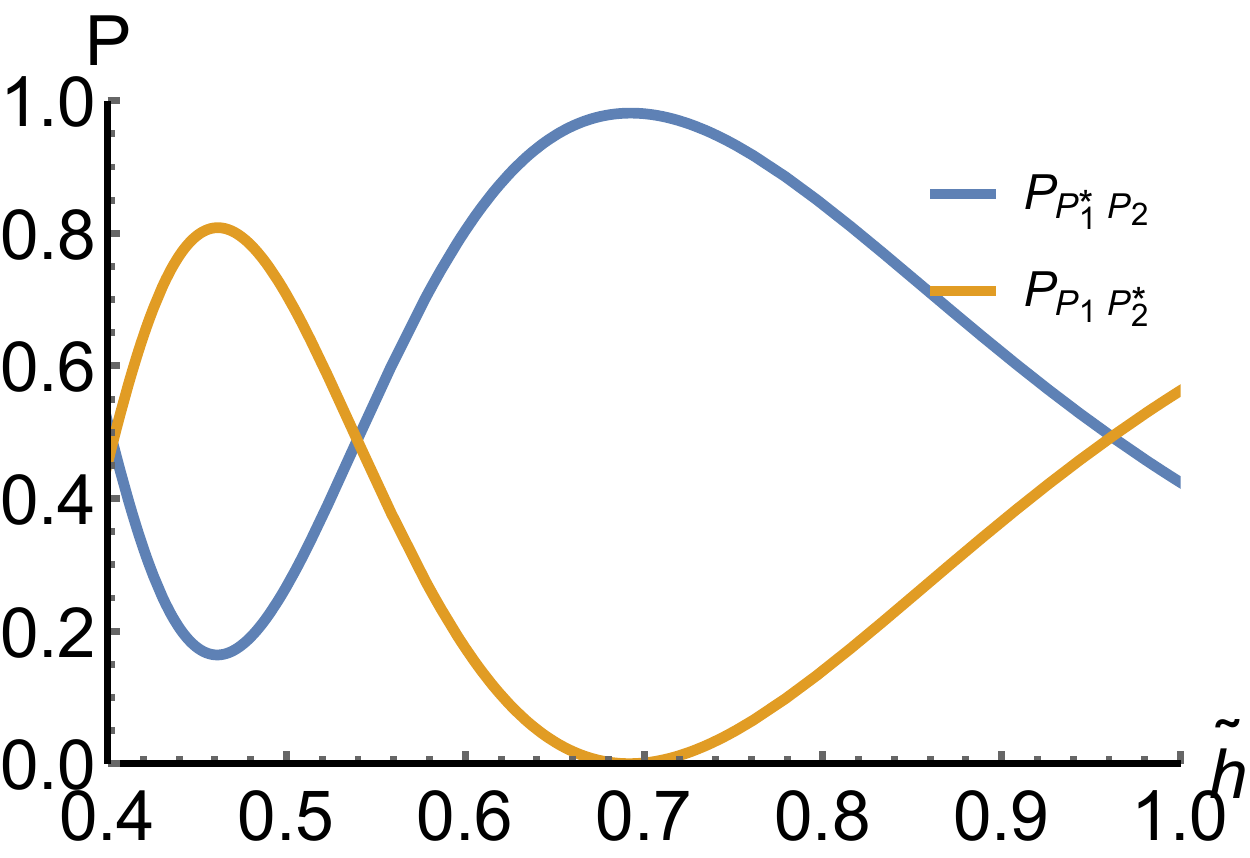}
 \caption{}
 \label{fig4(c)}
\end{subfigure}
\caption{Variation of the exciton site probabilities vs the macroscopic trait of the two-pigment system $P_{{P_1^*}{P_2}}$ (\textit{blue}) and $P_{{P_1}{P_2^*}}$ (\textit{orange}), (a) plot of $P_{{P_1^*}{P_2}}$ and $P_{{P_1}{P_2^*}}$ as a function of $\tilde{h}$ with the values $V=130$ cm$^{-1}$, $e-g=230$ cm$^{-1}$, and $\Gamma=0.01\Delta$ at a given time, (b) Same as (a) but with $V=230$ cm$^{-1}$, (c) Same as (a) but with $V=390$ cm$^{-1}$.}
\label{Fig4}
\end{figure*}
\section{The efficiency of the energy transfer in a two-pigment photosynthetic system}
\begin{figure*}[t!]
\begin{subfigure}{0.3\textwidth}
  \centering
 \includegraphics[trim=150 20 1 30, width=0.6\linewidth]{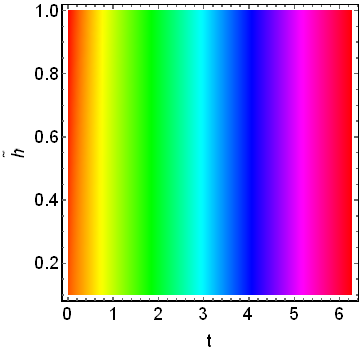}
 \caption{}
 \label{fig5(a)}
\end{subfigure}
~
\begin{subfigure}{0.3\textwidth}
  \centering
 \includegraphics[trim=150 20 1 30, width=0.6\linewidth]{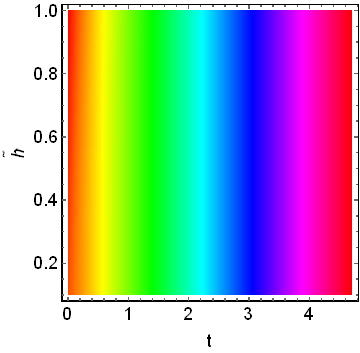}
 \caption{}
 \label{fig5(b)}
\end{subfigure}
~
\begin{subfigure}{0.3\textwidth}
  \centering
 \includegraphics[trim=150 20 1 30, width=0.6\linewidth]{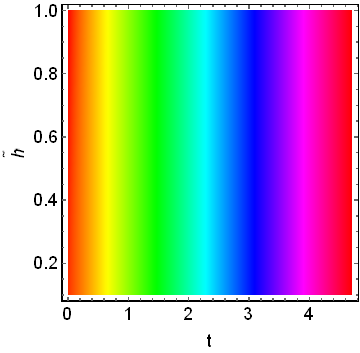}
 \caption{}
 \label{fig5(c)}
\end{subfigure}
~
\begin{subfigure}{0.0085\textwidth}
  \centering
 \includegraphics[trim=36 20 1 70, width=1.8\linewidth]{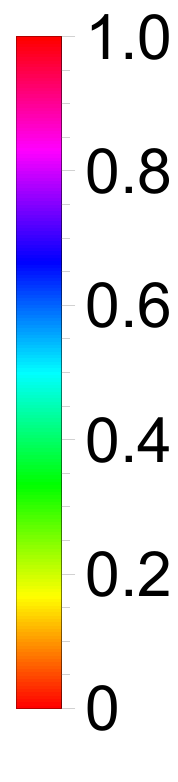}
 \label{fig5(d)}
\end{subfigure}
\caption{Density plot of quantum efficiency vs $\tilde{h}$ and $t$ with a fixed interaction energy values for each plot. (a) Plot of quantum efficiency with the values $V=130$ cm$^{-1}$, $e-g=230$ cm$^{-1}$ and $\Gamma=0.01\Delta$, (b) Same as (a) but with $V=230$ cm$^{-1}$, (c) Same as (a) but with $V=390$ cm$^{-1}$.}
\label{Fig5}
\end{figure*}

\begin{figure*}[t!]
\begin{subfigure}{0.33\textwidth}
  \centering
 \includegraphics[trim=10 40 10 1, width=1.3\linewidth]{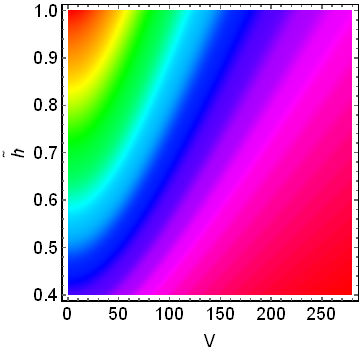}
 \label{fig6(a)}
\end{subfigure}
~
\begin{subfigure}{0.28\textwidth}
  \centering
 \includegraphics[trim=1 1 1 50, width=0.27\linewidth]{EfficContour-tCons-legend}
 \label{fig6(b)}
\end{subfigure}
\caption{Density plot of energy transfer efficiency vs $\tilde{h}$ and $V$ at a given time. The ratio $\tilde{h}/V$ can be considered as a prameter that governs the exciton transfer efficiency at a given time.}
\label{Fig6}
\end{figure*}

The exciton transfer process is often quantified employing an indicator, namely the energy transfer efficiency $\eta$ \cite{33, 36}. This quantity can be easily computed in terms of the time evolution of the one-exciton density matrix as
\begin{equation}\label{Eq22}
\eta(t)=2\sum_n R_n\int_0^t\langle e_n\vert \rho(t) \vert e_n \rangle dt
\end{equation}
The state of the system at time $t$ can be obtained using the completeness relation $\sum\vert e_n\rangle\langle e_n\vert=1$ and Eqs. (\ref{Eq10}) and (\ref{Eq11}), as follows
\begin{equation}
\label{Eq23}
\vert\Psi(t)\rangle\rangle=\sum_n\vert e_n\rangle\langle e_n\vert\hat{U}_I(t)\vert\Psi(0)\rangle\rangle=\sum_n\vert e_n\rangle\vert \widetilde{\chi_{e_n}(t)}\rangle
\end{equation}
Using the definition of density matrix $\rho(t)=\vert\Psi(t)\rangle\langle\Psi(t)\vert$ and by taking the trace over the environmental states, we obtain the density operator of the two-pigment system at time $t$ as $\rho_s(t)=P_{{P_1^*}{P_2}}\vert e_1\rangle\langle e_1\vert+P_{{P_1}{P_2^*}}\vert e_2\rangle\langle e_2\vert$. We can rewrite Eq. (\ref{Eq22}) in the form
\begin{align}\label{Eq24}
\eta(t) & =2 R_1\int_0^t \langle{{P_1^*}{P_2}}\vert\rho\vert{{P_1^*}{P_2}}\rangle dt +2 R_2\int_0^t \langle{{P_1^*}{P_2}}\vert\rho\vert{{P_1^*}{P_2}}\rangle dt \nonumber \\ &= 2 R_1\int_0^t P_{{P_1^*}{P_2}} dt + 2 R_2\int_0^t P_{{P_1}{P_2^*}} dt 
\end{align}
Substituing the expressions $P_{{P_1^*}{P_2}}$ and $P_{{P_1}{P_2^*}}$ from Eq. (\ref{Eq21}) and also inserting the parameters $\theta$ and $\Delta$ as defined in Eqs. (\ref{Eq5}) and (\ref{Eq6}), respectively, one can calculate the final expression for $\eta(t)$ as the Eq. (\ref{Eq25}). Here, the efficiency $\eta$ can be estimated in four different situations. First of all, we assume that the trapping rates at each pigment both are non-zero and have unequal values $R_2 \neq R_1\neq0$. Thus, we have
 
\begin{align}\label{Eq25}
\eta(t) & =2R_1\lbrace
t.\cos^4\theta-\dfrac{1}{\Gamma}\sin^4\theta e^{-\Gamma t}+{2}\cos^2\theta\sin^2\theta e^{-\Gamma t/2}(\dfrac{-2\Gamma\cos \Delta t+4\Delta\sin \Delta t}{4\Delta^2+\Gamma^2})\rbrace\nonumber \\ 
+& 2R_2 \lbrace t.\cos^2\theta\sin^2\theta-\dfrac{1}{\Gamma}\cos^2\theta\sin^2\theta e^{-\Gamma t}-{2}\cos^2\theta\sin^2\theta e^{-\Gamma t/2}(\dfrac{-2\Gamma\cos \Delta t+4\Delta\sin \Delta t}{4\Delta^2+\Gamma^2})\rbrace
\end{align}
Second, we can calculate the quantum efficiency $\eta$ supposing that only trapping rate of the pigment in site $1$ has a non-zero value, i.e., $R_1\neq R_2=0$. So one gets
\begin{align}\label{Eq26}
\eta(t) & =2R_1\lbrace
t.\cos^4\theta-\dfrac{1}{\Gamma}\sin^4\theta e^{-\Gamma t}+{2}\cos^2\theta\sin^2\theta e^{-\Gamma t/2}(\dfrac{-2\Gamma\cos \Delta t+4\Delta\sin \Delta t}{4\Delta^2+\Gamma^2})\rbrace
\end{align}
Third, in the situation in which $R_2\neq R_1=0$, we have
\begin{align}\label{Eq27}
\eta(t) & = 2R_2 \lbrace t.\cos^2\theta\sin^2\theta-\dfrac{1}{\Gamma}\cos^2\theta\sin^2\theta e^{-\Gamma t}-{2}\cos^2\theta\sin^2\theta e^{-\Gamma t/2}(\dfrac{-2\Gamma\cos \Delta t+4\Delta\sin \Delta t}{4\Delta^2+\Gamma^2})\rbrace
\end{align}
And finally, if both trapping rates of each site have equal values $R_2=R_1=R$, one obtains
\begin{align}\label{Eq28}
\eta(t) & =2R\lbrace
t.\cos^2\theta-\dfrac{1}{\Gamma}\sin^2\theta e^{-\Gamma t}\rbrace
\end{align}
Fig. (\ref{Fig5}) shows the density plot of quantum efficiency $\eta(t)$ as a function of $\tilde{h}$ and $t$. Figs. (\ref{fig5(a)}) to (\ref{fig5(c)}) show the variation of efficiency $\eta(t)$ in simillar condition except than the interaction energy $V$ takes different value in each case. According to Fig. (\ref{Fig5}), we can see that at a fixed value of $V$, variation in macroscopic behavior of the supposed system may not alter the quantum efficiency $\eta(t)$. Therefore we find that the in this condition, the quantum efficiensy is robust with respect to the macroscopicity parameter $\tilde{h}$. \\Finally, Fig. (\ref{Fig6}) shows the plot of efficiency as a function of $\tilde{h}$ and the interaction energy $V$. This figure shows that at a given time, how the magnitude of $\tilde{h}$ may be important to reach an optimal region of exciton transfer efficiency. According to this figure we can consider the ratio $\tilde{h}/V$ as a prameter that governs the exciton transfer efficiency at a given time. At a high ratio of $\tilde{h}/V$ the photosynthetic system lies in a low efficient energy transfer region and as this ratio increases an optimal region of energy transfer efficiency emerges according to the Fig. (\ref{Fig6}).
 
\section{Conclusion}
In this article, we studied the problem of electronic energy transfer in a two-pigment system coupled to a harmonic environment using a semi-classical non-master equation formalism. We considered a two-dimensional Hamiltonian for a two-pigment photosynthetic organism, involving trapping terms, to investigate the probability of the exciton transfer between two pigment states in the presence of an oscillating environment. We used the second-order perturbation theory to calculate the time-dependent populations of each excitonic state in photosynthetic system. Results illustrated that Our results demonstrate that the quantum efficiency is robust concerning the macroscopicity parameter $\tilde{h}$ individually, but the ratio of macroscopicity over the pigment-pigment interaction energy $V$ can be considered as a parameter that may govern the quantum efficiency at a given time. So, the dynamical behavior and the quantum efficiency for transport phenomena in photosynthetic systems may be influenced by the macroscopic quantum trait of the system. We can conclude that the degree of the macroscopic behavior of the photosynthetic system has a significant role in the dynamics of the energy transfer in these systems. Although, the scientific significance of photosynthesis is indisputable, achieving a highly efficient exciton transport model in light-harvesting complexes will be honored from the technological perspective, too. A precise knowledge of coherent dynamics for energy transfer in photosynthetic organisms is speculated to alter the microscopic view of energy transfer in both physical and biological systems in the future. As a new achievement our result may be significant to design synthetic devices for transportation based on macroscopic quantum phenomena.
\\

\end{document}